\input{aipcheck}

\documentclass[
    ,final            % use final for the camera ready runs
  ]
  {aipproc}

\layoutstyle{6x9}

\def\lsim{\raise0.3ex\hbox{$\;<$\kern-0.75em\raise-1.1ex
\hbox{$\sim\;$}}}
\def\gsim{\raise0.3ex\hbox{$\;>$\kern-0.75em\raise-1.1ex
\hbox{$\sim\;$}}}

\usepackage[latin1]{inputenc}
\usepackage{pstricks,pst-node,pst-text,pst-3d}
\usepackage{amsmath}
\usepackage{tabularx}

\begin{document}
\title{
\vspace{-1.6cm}
\hglue 12cm {\small IFT-P.025/2003} \\
\vspace{0.5cm}
CP and T Violation in Neutrino Oscillations
\footnote{Invited talk given by H. Nunokawa at 
the 10 th Mexican School of Particles and Fields 
at Playa del Carmen, Mexico, Oct. 30 - Nov. 6, 2002}}

\author{
\vspace{-0.3cm}
Hisakazu Minakata$^{\ a}$, Hiroshi Nunokawa$^b$ and Stephen Parke$^c$}{
  address={
\vspace{-0.2cm}
$^a$Department of Physics, Tokyo Metropolitan University \\
1-1 Minami-Osawa, Hachioji, Tokyo 192-0397, Japan\\[0.16cm]
$^{\ b}$
Instituto de F{\'\i}sica Te{\'o}rica,Universidade Estadual Paulista, \\
    Rua Pamplona 145, 01405-900 S{\~a}o Paulo, Brazil\\[0.16cm]
$^{\ c}$
Theoretical Physics Department,
Fermi National Accelerator Laboratory \\
P.O.Box 500, Batavia, IL 60510, USA
}
}

\begin{abstract}
In this short lecture, we discuss some basic phenomenological 
aspects of CP and T violation in neutrino oscillation. 
Using CP/T trajectory diagrams in the bi-probability space, 
we try to sketch out some essential features of 
the interplay between the effect of CP/T violating phase 
and that of the matter in neutrino oscillation. 
\end{abstract}

\maketitle

%%%%%%%%%%%%%%%%%%%%%%%%%%%%%%%%%%%%%%%%%%%%
%% MAINMATTER
%%%%%%%%%%%%%%%%%%%%%%%%%%%%%%%%%%%%%%%%%%%%
\vspace{-1.0cm}
\section{Introduction}
There is now evidence for neutrino oscillations coming 
from the observation of atmospheric neutrinos~\cite{atmnu}, 
solar neutrinos~\cite{solar}, as well as neutrinos produced
by accelerator~\cite{k2k}.
In particular, recent evidence for 
the disappearance of $\bar{\nu}_e$ coming from nuclear 
reactors reported by the KamLAND experiment~\cite{kamland}
has finally established the 
so called large mixing angle (LMA)
MSW~\cite{msw} solution to the solar neutrino problem
and has opened the door to explore 
the CP/T violating phase $\delta$ in the 
Maki-Nakagawa-Sakata (MNS)~\cite{mns} neutrino 
mixing matrix through neutrino oscillation~\cite{cpv}.
In this talk we will discuss some basic aspects of CP 
and T violation in neutrino oscillation 
using bi-probability trajectories diagrams,
which are quite useful for qualitative 
understanding of the subject. 
\vspace{-0.5cm}
\section{Three Neutrino Flavor Mixing Scheme}

Let us consider the neutrino mixing among three flavor as
\vglue -0.4cm
\begin{equation}
\nu_\alpha = \sum_{i=1}^{3} U_{\alpha i} \nu_i 
\end{equation}
\vglue -0.05cm
\noindent
where $\nu_\alpha (\alpha=e,\mu,\tau)$ and
$\nu_i (i=1,2,3)$ are weak and mass eigenstates, respectively, 
and $U$ is the MNS~\cite{mns} 
neutrino mixing matrix. 
We will adopt the standard parametrization~\cite{PDG02} 
of the MNS matrix as follows,  
\vglue -0.1cm
\begin{equation}
U =
\left[
\begin{matrix}
1  &  0       & 0       \cr 
0  &  c_{23}  & s_{23}  \cr 
0  & -s_{23}  & c_{23}  
\end{matrix}
\right]
\left[
\begin{matrix}
c_{13} & 0  &  s_{13} \text{e}^{-i\delta}\cr 
0  & 1 & 0 \cr 
-s_{13}\text{e}^{i\delta}  & 0 & c_{13} 
\end{matrix}
\right]
\left[
\begin{matrix}
 c_{12} & s_{12} & 0 \cr 
-s_{12} & c_{12} & 0 \cr 
0  & 0 & 1 
\end{matrix}
\right], 
\label{eq:mns_pdg}
\end{equation}
\vglue 0.05cm
\noindent
where $c_{ij} \equiv \cos\theta_{ij}$, 
$s_{ij} \equiv \sin\theta_{ij}$ and
$\delta$ is the CP violating phase.  

We already have significant amount of information about 
the MNS matrix as well as neutrino mass squared 
differences ($\Delta m^2_{ij} \equiv m^2_{j}- m^2_{i}$)
from various experiments. 
See, e.g., Ref.~\cite{review}, for a recent review.
Under the parametrization in Eq.~(\ref{eq:mns_pdg}),
the atmospheric neutrino data indicate~\cite{atmnu}
\vglue -0.3cm
\begin{equation}
\sin^2 2\theta_{23} \simeq 0.88-1, \ \ 
|\Delta m^2_{23}| \simeq |\Delta m^2_{13}| \simeq 
(1.5-4) \times 10^{-3} \text{eV}^2,
\end{equation}
whereas the solar neutrino data combined with 
the KamLAND one indicate~\cite{sol-kam},  
\begin{equation}
\tan^2 2\theta_{12} \simeq 0.25-0.85, \ \ 
\Delta m^2_{12} \simeq (4-20) \times 10^{-5} \text{eV}^2.
\end{equation}
On the other hand, negative result of CHOOZ experiment
impose constraint on $\theta_{13}$ as~\cite{chooz},   
\begin{equation}
\sin^2 2\theta_{13} \lsim 0.1
\end{equation}

However, toward the complete understanding
of the neutrino sector, there still remain 
following three questions to be answered, 
1) what is the sign of $\Delta m^2_{13}$?, \ 
normal ($\Delta m^2_{13}>0$) or inverted ($\Delta m^2_{13}<0$)? 
2) how small is the value of $\theta_{13}$?, \ 
3) what is the value of CP phase $\delta$?
Hereafter, we mainly focus on the effect of the 
CP phase $\delta$ in neutrino oscillation, which can 
not be separately considered from that of 
the sign of $\Delta m^2_{13}$ as well as 
the value of $\theta_{13}$, as we will see. 

\section{CP and T Violation in Neutrino Oscillation}

Let us first consider the effect of CP and T violation
in neutrino oscillation in vacuum~\cite{cpv}. 
If the oscillation probability of 
$\nu_{\beta}\rightarrow\nu_{\alpha}$ is different from 
its CP conjugate process, 
$\bar{\nu}_{\beta}\rightarrow\bar{\nu}_{\alpha}$, 
or if
\vglue -0.2cm
\begin{equation}
\Delta P^{CP}_{\alpha \beta} \equiv
P(\nu_{\beta}\rightarrow\nu_{\alpha})-
P(\bar{\nu}_{\beta}\rightarrow\bar{\nu}_{\alpha})
\ne 0
\hskip 0.5cm 
(\alpha, \beta = e,\mu, \tau, \alpha \ne \beta),
\label{eq:delta_cp}
\end{equation} 
for a given neutrino energy ($E$) and baseline ($L$), 
then this implies CP violation. 
Similarly,
if the oscillation probability of 
$\nu_{\beta}\rightarrow\nu_{\alpha}$ is different from 
its T conjugate (time reversal) process, 
${\nu}_{\alpha}\rightarrow \nu_{\beta}$, 
or 
\begin{equation}
\Delta P^{T}_{\alpha \beta} \equiv
P(\nu_{\beta}\rightarrow\nu_{\alpha})-
P(\nu_{\alpha}\rightarrow \nu_\beta)\ne 0
\hskip 0.5cm 
(\alpha, \beta = e,\mu, \tau, \alpha \ne \beta),
\label{eq:delta_t}
\end{equation} 
then this implies T violation. 
If the CPT symmetry holds, which is the case for
neutrino oscillation in vacuum, 
violation of T is equivalent to that of CP. 

Using the parametrizaion in Eq. (\ref{eq:mns_pdg}), 
one can explicitly show that in vacuum
$\Delta P^{CP}_{\alpha \beta}$ and 
$\Delta P^{T}_{\alpha \beta}$ defined 
in Eqs. (\ref{eq:delta_cp}) and (\ref{eq:delta_t}) are
equal and given by, 
\vglue -0.2cm
\begin{equation}
\Delta P^{CP}_{\alpha \beta} = \Delta P^{T}_{\alpha \beta}
= -16J_{\beta\alpha}
\sin\left(\frac{\Delta m_{12}^2}{4E}L\right)
\sin\left(\frac{\Delta m_{23}^2}{4E}L\right)
\sin\left(\frac{\Delta m_{13}^2}{4E}L\right),
\label{eq:deltaP}
\end{equation}
where 
\vglue -0.2cm
\begin{equation}
J_{\beta\alpha}\equiv \mbox{Im}
[U_{\alpha 1}U_{\alpha 2}^*U_{\beta 1}^*U_{\beta 2}]
= \pm c_{12}s_{12}c_{23}s_{23}c_{13}^2s_{13}\sin\delta,
\end{equation}
with $+(-)$ sign is for cyclic (anti-cyclic) permutation 
of $(\alpha,\beta)=(e,\mu),(\mu,\tau),(\tau,e)$.
Note that in order that the CP/T violation effect to be
non-zero, all the angle must be non-zero and therefore, 
three flavor mixing is essential 
(no CP/T violation in two generation). 

One can estimate the effect of CP/T violation in vacuum 
using the best fitted values of the mixing parameters 
obtained from solar and atmospheric neutrino data as, 
\vglue -0.3cm
\begin{equation}
\displaystyle
\Delta P^{CP}_{\alpha \beta} = \Delta P^{T}_{\alpha \beta} 
\sim 2
\left[ \frac{\sin^2 2\theta_{12}}{0.83}\right]
^{\frac{1}{2}}
\left[ \frac{\Delta m^2_{12}}{7\times 10^{-5}\ \text{eV}^2}
\right]
\left[ \frac{\sin^2 2 \theta_{13}}{0.05}\right]^{\frac{1}{2}}
\sin \delta \ \ [\%],
\end{equation}
where we assumed that oscillation probability is measured
at energy and baseline when  $\Delta m^2_{13}L/4E $ takes 
$\pi/2$ (oscillation maximal). 
The CP/T violation effect is expected to be a few percent provided 
that $\theta_{13}$ is close to the CHOOZ limit, 
and can in principle be measurable by the proposed 
long-baseline neutrino oscillation experiments~\cite{future-lbl}
in the near future. 
Let us note that if the solution to the solar neutrino 
problem were not LMA but some another one which requires much 
smaller $\Delta m^2_{12}$ or $\theta_{12}$,
then the measurement of CP/T violation effect 
would be much more difficult!
\vglue -0.2cm
\section{CP and T trajectories in bi-probability space}

Let us now consider the effect of matter. In mater, 
measurement of CP violation can become more complicated 
because of the fact that oscillation probability 
for neutrinos and anti-neutrinos are in general different
in matter even if $\delta = 0$. 
Indeed, the matter effect can either contaminate or 
enhance the effect of intrinsic CP violation
effect coming from $\delta$~\cite{cpvmatter}.
For the case of T violation, the situation is different. 
If we can establish $\Delta P^{T}_{\alpha \beta}\ne0$ for
$\alpha\ne\beta$, then this imply $\delta \ne 0$
even in the presence of matter. This is because 
oscillation probability is invariant under time 
reversal even in the presence of matter.
Similar to the case of CP violation, 
T violation effect can either be enhanced 
or suppressed in matter~\cite{PW}.
However, T violation measurement is experimentally 
more difficult to perform, because we need to make
a non-muon neutrino beam!

Let us try to look into more about the interplay 
between the CP/T violation 
effect and matter effect. In order to have more transparent 
understanding of the subject, 
We will introduce the CP and T trajectory diagrams 
in the bi-probability space which were suggested 
and developed in Refs.~\cite{MNjhep01,mnp1,mnp2}.
From now on, we will focus on oscillation only for 
$\nu_\mu \leftrightarrow \nu_e$ as well as 
$\bar{\nu}_\mu \leftrightarrow \bar{\nu}_e$ channels 
which are experimentally more feasible, 
because the production and detection of 
${\nu}_\tau/\bar{\nu}_\tau$ are 
much more difficult. 

In vacuum one can show very easily that oscillation 
probabilities for neutrino and anti-neutrino channels 
take, without any approximation, the following forms, 
\vspace{-0.1cm}
\begin{eqnarray}
P &\equiv& 
P(\nu_\mu \rightarrow \nu_e) = A \cos \delta + B \sin \delta +C, 
\nonumber \\
\bar{P} &\equiv& P(\bar{\nu}_\mu \rightarrow \bar{\nu}_e) 
= A \cos \delta - B \sin \delta +C, 
\label{eq:prob}
\end{eqnarray}
\vglue -0.2cm
\noindent
where $A$, $B$ and $C$ are some constant 
which depend on mixing parameters, 
$\theta_{ij}$ and $\Delta m^2_{ij}$
as well as neutrino energy and baseline. 
Suppose that energy and baseline are fixed to some values. 
Then a given value of $\delta$ (e.g. $\delta$ = 0) defines 
one point $(P_0,\bar{P}_0)$ in the $P-\bar{P}$ plane. 
If we vary $\delta$ from $0$ to $2\pi$ we can 
draw a closed trajectory, which is an ellipse, 
in the $P-\bar{P}$ plane. 
This is schematically illustrated in Fig.~1.
%%%%%%%%%%%%%%%%%%%%%%%%%%%%%%%%%%%%%%%%%%%%%%%%%%%%%%%%%%%%
\vglue 0.3cm
\hglue -0.5cm
\includegraphics[height=.27\textheight]{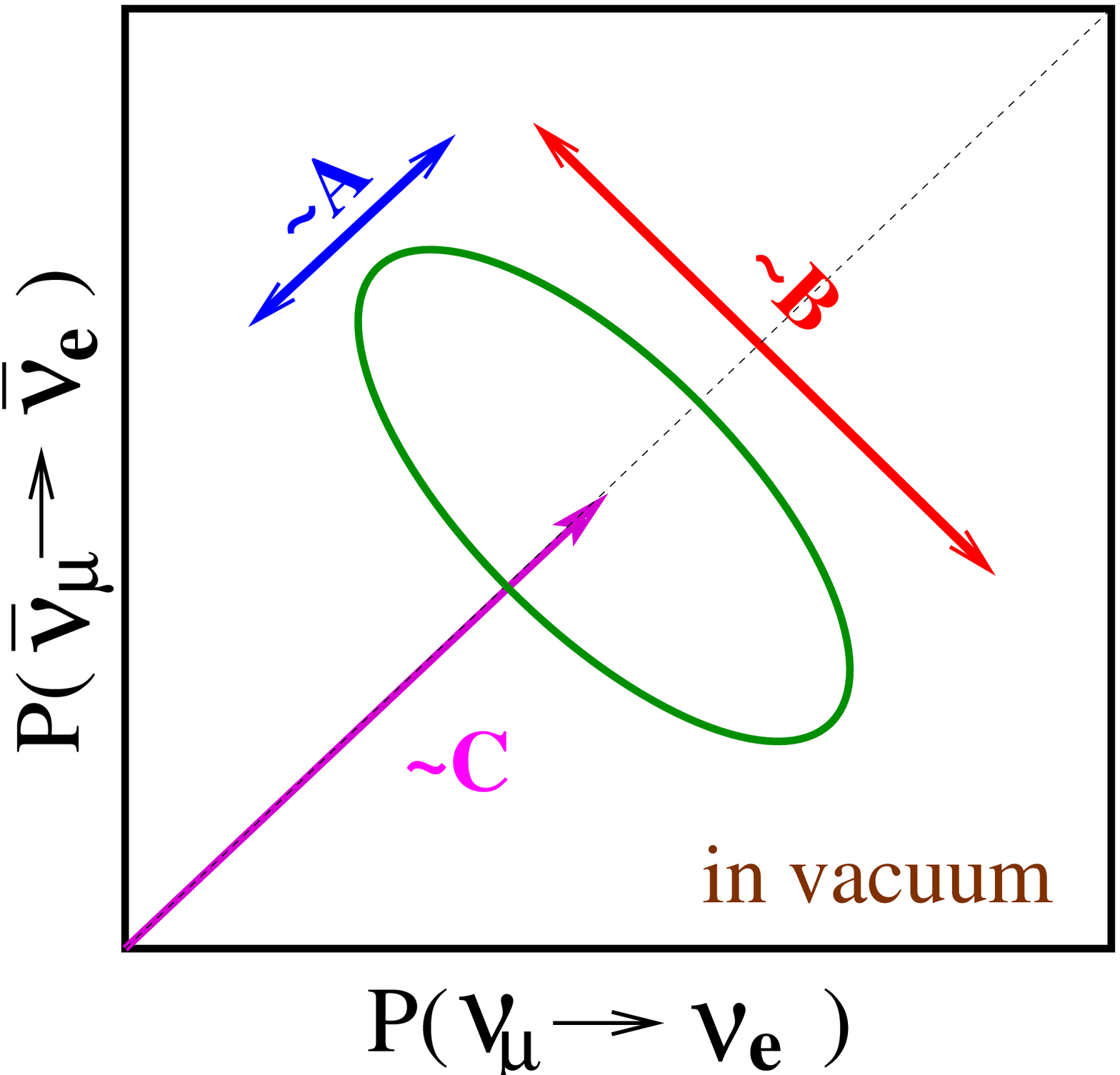}
\vglue 0.1cm
\hglue -0.4cm
\parbox[]{6.3cm}
{\small Fig. 1: Schematic illustration of the 
CP trajectory diagram without matter 
effect in the bi-probability $P-\bar{P}$ plane.}
\vglue -7.8cm 
\hglue 6.2cm 
\parbox[]{8.2cm}
{
\ \ There are several important features to be mentioned. 
First the constant $C$,\ which is 
proportional to $\sin^2\theta_{13}$, 
determines how far the ellipse is 
located from the origin of the the $P-\bar{P}$ plane. 
Second, the size of the ellipse, which is determined 
by the magnitudes of $A$ and $B$, corresponds 
to the size of the effect of non-zero CP phase
in oscillation probabilities. 
To be more precise, the size of the ellipse along 
the axis which is proportional to $B$ characterize 
the size of direct CP violation effect 
which is proportional to $P-\bar{P} \propto \sin \delta$ 
whereas that of the direction along the axis 
proportional to $A$
characterize the effect of CP ``conserving'' term,
which is proportional to $\cos \delta$.
We note that the minor (if 
$A > B$) or major (if $A < B$) 
axis is always at 45 degree. 
On the \break
}
{\vglue -0.8cm}
\noindent
trajectory, the two special points which sit 
at $P = \bar{P}$ correspond to cases 
where $\delta$ takes either 0 or $\pi$. 
Note that when $\delta=0$ or $\pi$, the mixing
matrix become real and no difference between 
neutrino and anti-neutrino oscillation probability.

How the matter effect can change this picture?
It has been noted in Ref.~\cite{KTY02} that even 
in the matter with arbitrary density profile, 
the oscillation probability can be 
expressed \break
%%%%%%%%%%%%%%%%%%%%%%%%%%%%%%%%%%%%%%%%%%%%%%%%%%%%%%%%%%%%
\vglue -0.1cm
\hglue -0.5cm
\includegraphics[height=.27\textheight]{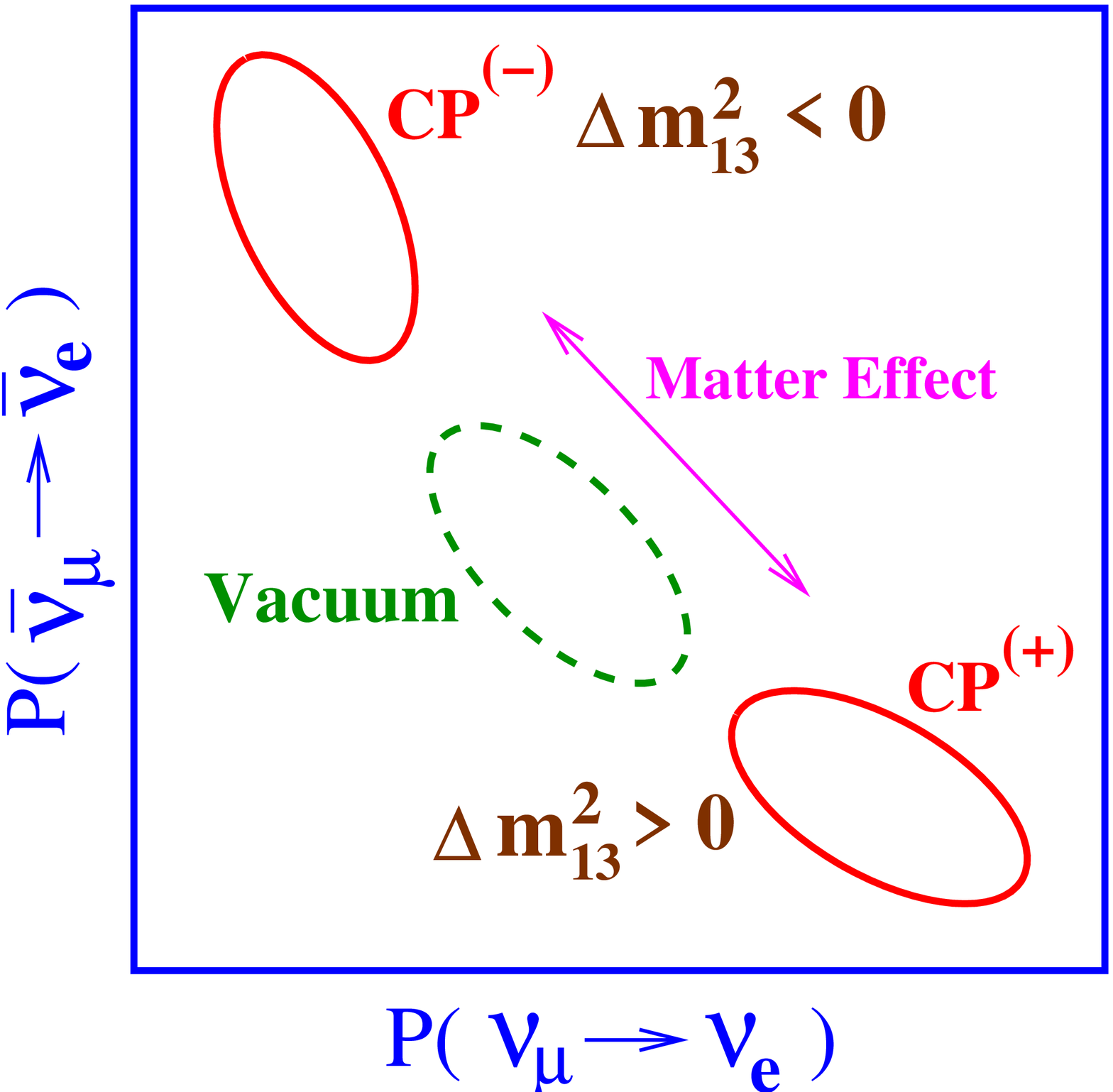}
\vglue 0.2cm
\hglue -0.4cm
\parbox[]{6.3cm}
{\small Fig. 2: Schematic illustration of how 
the trajectory in the $P-\bar{P}$ plane is
affected by the matter effect. 
}
\vglue -7.9cm 
\hglue 6.2cm 
\parbox[]{8.2cm}
{
in the same form as in vacuum in Eq.~(\ref{eq:prob}), 
without any approximation, as 
\begin{equation}
P = \tilde{A} \cos \delta + \tilde{B} \sin \delta +\tilde{C},
\end{equation}
where $\tilde{A}$, $\tilde{B}$ and $\tilde{C}$ are some 
constant which depend not only on mixing angle but also 
on the matter effect and are different for neutrinos
and anti-neutrinos.
This implies that the trajectory in matter is also 
elliptic but is 
shifted to two different directions, 
according to the sign of 
$\Delta m^2_{13}$, 
in the $P-\bar{P}$ plane
as illustrated in Fig. 2. 
\ \ What happens with matter effect is that the size of 
the trajectory does not change essentially 
but change
its position in the $P-\bar{P}$ plane
due to %{\hglue 0.8cm}
some parallel shift plus rotation, 
as illustrated in Fig. 2. For 
positive (negative) value of $\Delta m^2_{13}$,
due to the matter effect, 
probability for 
\break
} 
%%%%%%%%%%%%%%%%%%%%%%%%%%%%%%%%%%%%%%%%%%%%%%%%%%%%%%%%%%%%
\vglue -0.3cm
\noindent
$\nu_{\mu}\rightarrow\nu_e $ 
($\bar{\nu}_{\mu}\rightarrow\bar{\nu}_e $) is
enhanced whereas that for 
$\bar{\nu}_{\mu}\rightarrow\bar{\nu}_e $ 
($\nu_{\mu}\rightarrow\nu_e $ ) is suppressed
so that they are sifted as in Fig. 2.
The magnitude of shift is larger when the 
the matter effect is larger, i.e., 
the baseline is larger. 
In fact, 
if the distance is short the two trajectories labeled as 
CP$^{(+)}$ and CP$^{(-)}$ can be overlapped
(see Fig. 5 for an explicit example for such a case).
We note that trajectories labeled as 
CP$^{(+)}$ and CP$^{(-)}$ are symmetric with 
respect to $P=\bar{P}$ line, which we will
explain why later.

Let us next consider the T violation. We can do 
exactly the same exercise as we did 
%%%%%%%%%%%%%%%%%%%%%%%%%%%%%%%%%%%%%%%%%%%%%%%%%%%%%%%%%%%%
\vglue 0.2cm
\hglue -0.2cm
\includegraphics[height=.29\textheight]{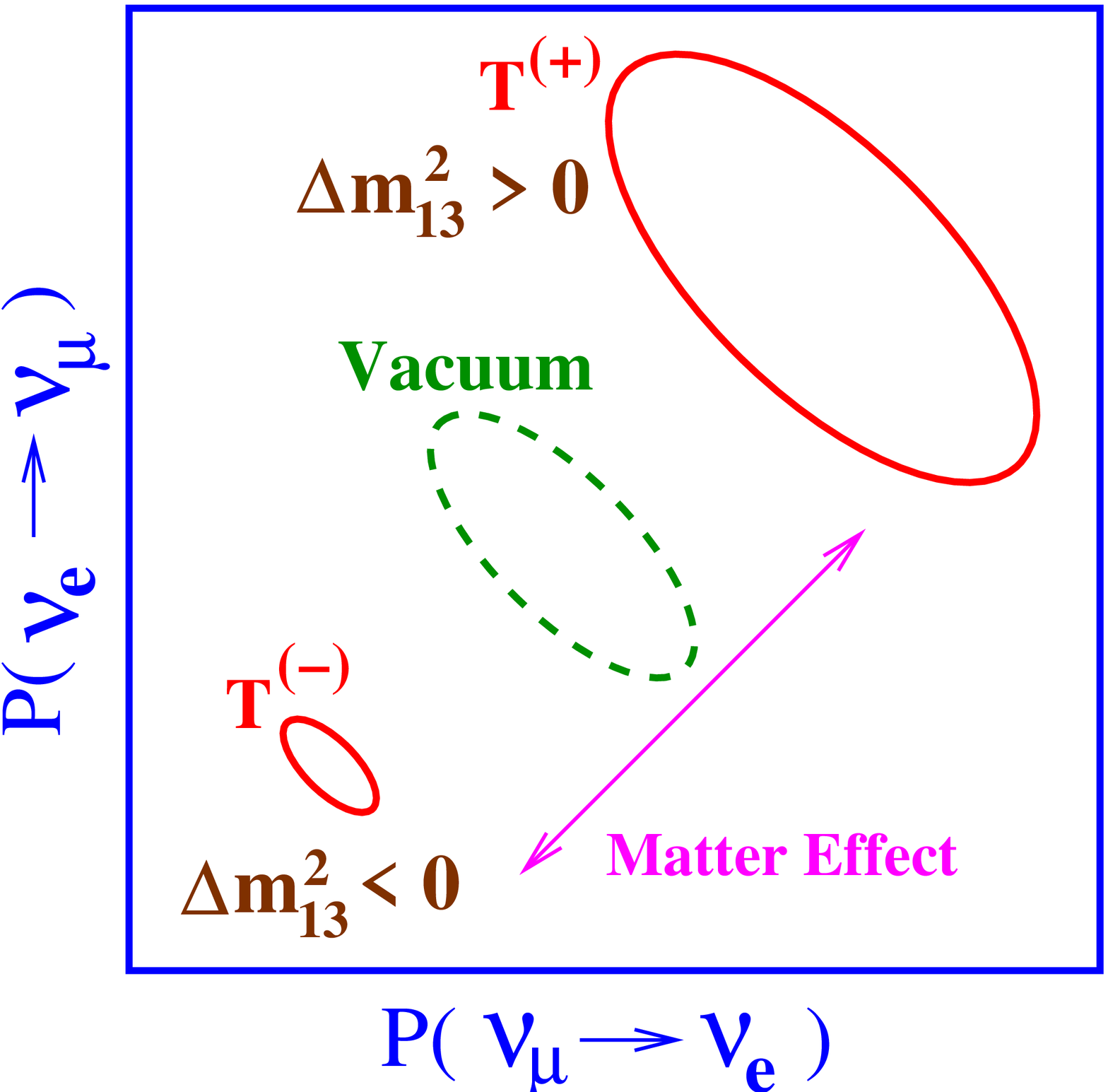}
\vglue 0.2cm
\hglue -0.2cm
\parbox[]{6.4cm}
{
\small 
Fig. 3: Schematic illustration of the 
trajectory diagram in the bi-probability 
plane $P-P^T$.
}
\vglue -8.4cm 
\hglue 7.0cm 
\parbox[]{7.3cm}
{in Fig. 2
but in the bi-probability $P-P^T$ plane where 
$P^T\equiv P(\nu_e\to\nu_\mu)$ is probability 
for the 
T conjugate channel of the $\nu_\mu\to\nu_e$ process.
In contrast to the case of CP trajectory diagram, 
not only the position but also the size of 
the ellipse changes in matter as illustrated in 
Fig. 3. 
One can easily understand the qualitative 
behaviour of these trajectories 
by noting that 
due to the matter effect both 
$P$ and $P^T$ are enhanced (suppressed) 
for positive (negative) $\Delta m^2_{13}$~\cite{PW}.
The magnitude of the shift as well as 
the size change of the ellipse depend on 
the strength of the matter effect. 
Larger the matter effect (longer the baseline), 
larger the sift and the size change.
}
%%%%%%%%%%%%%%%%%%%%%%%%%%%%%%%%%%%%%%%%%%%%%%%%%%%%%%%%%%%%
\vglue -0.4cm
\section{Unified understanding of CP and T trajectories}
\vglue -0.2cm
Now we will try to relate these two kind of different
trajectories which represent 
CP and \break
%%%%%%%%%%%%%%%%%%%%%%%%%%%%%%%%%%%%%%%%%%%%%%%%%%%%%%%%%%%%
\vglue -0.2cm
\hglue -0.6cm
\includegraphics[height=.32\textheight]{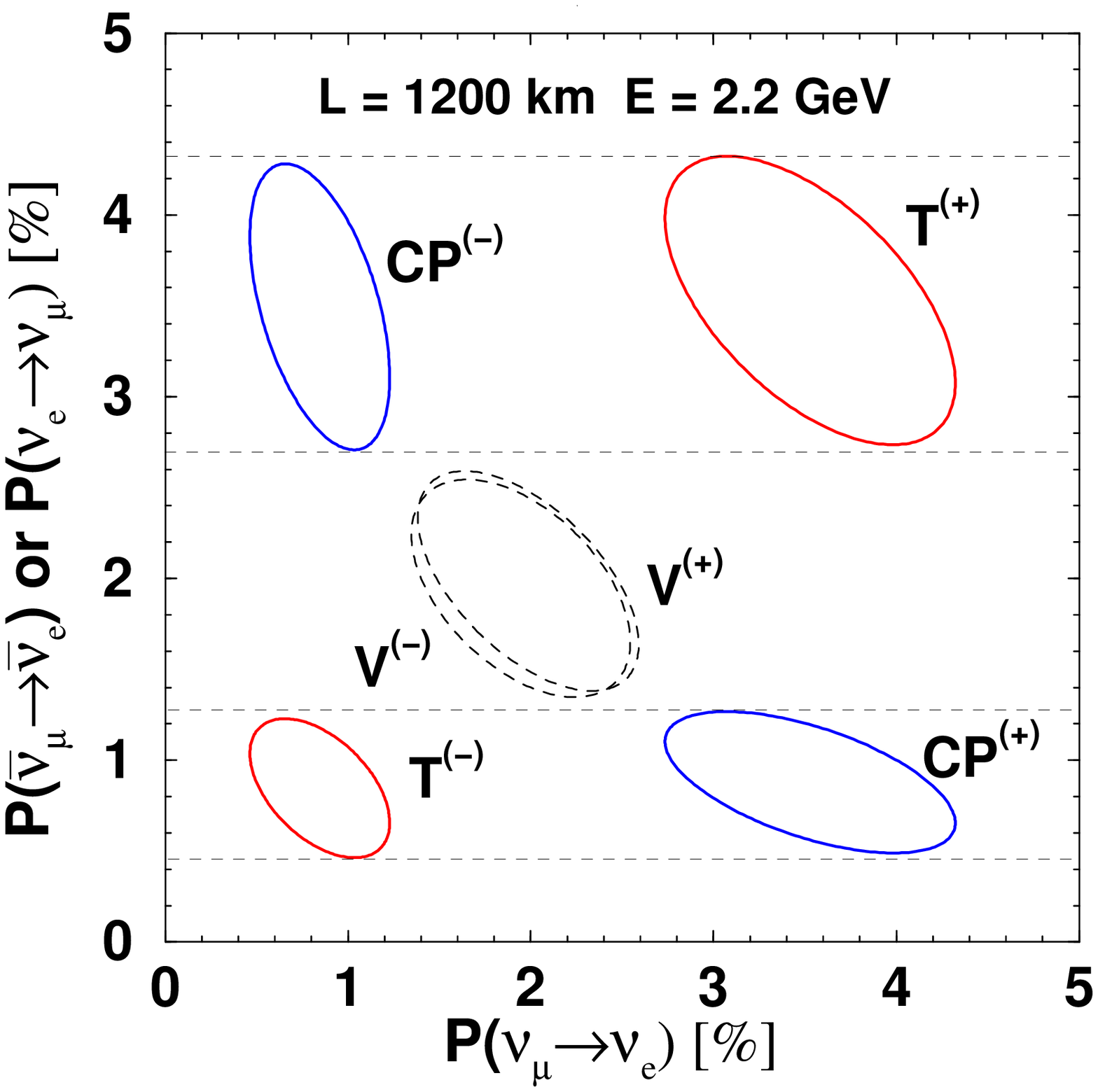}
\vglue 0.1cm 
\parbox[]{6.2cm}
{
\small 
Fig. 4: 
Relation between CP and T trajectory diagrams.
}
\vglue -8.4cm 
\hglue 6.9cm 
\parbox[]{7.5cm}
{T violations in 
the presence of matter and give a unified picture~\cite{mnp1}. 
In Fig. 4, we draw 
the CP and T trajectories in matter at the same
time in $P-\bar{P}$ or $P-P^T$ plane
together with that in vacuum (labeled as $V^{(\pm)}$). 
For this plot, we set 
$E=2.2$ GeV, $L$=1200 km, 
and 
$|\Delta m^2_{13}| = 3 \times 10^{-3}$ eV$^2$,
$\sin^2 2\theta_{23}=1.0$,
$\Delta m^2_{12} = +5 \times 10^{-5}$ eV$^2$,
$\sin^2 2\theta_{12}=0.80$, $\sin^2 \theta_{13}=0.05$
and the electron density $ Y_e \rho  = 1.5$ g cm$^{-3}$.
The superscript $(\pm)$ attached to the label CP/T/V 
indicate the sign of $\Delta m^2_{13}$.
We note that two $V^{(\pm)}$ trajectories are almost degenerated
(We ignored the difference between $V^{(+)}$ and $V^{(-)}$ 
in the previous sections in Figs. 1-3
for simplicity).
The vertical axis must be identified with 
$\bar{P}$ and $P^T$ for CP and T trajectories, respectively. 
There is a remark-\break
\break
}
\vglue -0.8cm
\noindent
able feature.
Projections of CP$^{(\mp)}$ and T$^{(\pm)}$ 
trajectories to the vertical axis ``almost''
(but not exactly) coincide with 
each other. 
Let us try to explain why.  
\vglue -0.8cm
\newpage
In the presence of matter,
one must consider the neutrino evolution 
equation given as,
\vglue -0.3cm
\begin{equation}
i\frac{d}{dt}
\left[
\begin{matrix} 
\nu_e \cr \nu_\mu \cr \nu_\tau \cr
\end{matrix} 
\right] 
= 
\left[ 
U \mbox{diag} (0, \frac{\Delta m^2_{12}}{2E},
\frac{\Delta m^2_{13}}{2E}) U^\dagger
+ \mbox{diag} (a,0,0) \right] 
\left[
\begin{matrix} 
 \nu_e \cr \nu_\mu \cr \nu_\tau \cr
\end{matrix} 
\right], 
\label{eq:evol}
\end{equation}
where $a = \sqrt{2} G_F N_e$ denotes the index of refraction 
for $\nu_e$ in matter with $G_F$ being the Fermi constant 
and $N_e$ a constant electron number density in the earth. 
For anti-neutrinos, the same equation holds but with 
$a\to -a$ and $U\to U^\star$.

By taking the complex conjugate of this evolution equation, 
one can show that the equation for neutrino is
identical to that of anti-neutrinos with the sign of $\Delta m^2$'s
flipped, from which we can conclude that 
for arbitrary matter density profile, 
\vspace{-0.1cm}
\begin{eqnarray}
P(\nu_{\mu} \rightarrow \nu_{e};
\Delta m^2_{13},\Delta m^2_{12}, \delta,a)
&=&
P(\bar{\nu}_{\mu} \rightarrow \bar{\nu}_{e};
-\Delta m^2_{13},-\Delta m^2_{12},\delta, a).
\label{eq:CP-CP-exact}
\end{eqnarray}
We call this CP-CP relation, which holds without any 
approximation.

On the other hand, by taking the time reversal 
($t\to -t$) of the evolution equation, one can show
that the equation for neutrino which 
describe $\nu_e \to\nu_\mu$ process is 
identical to that for its T conjugate process 
but for anti-neutrinos $\bar{\nu}_\mu \to\bar{\nu}_e$ 
with the signs of $\Delta m^2$'s and $\delta$ 
flipped, from which we can conclude that 
without any approximation,
\vspace{-0.1cm}
\begin{eqnarray}
P(\nu_e \rightarrow \nu_{\mu};
\Delta m^2_{13},\Delta m^2_{12}, \delta, a)
&=&
P(\bar{\nu}_{\mu} \rightarrow \bar{\nu}_{e};
-\Delta m^2_{13},-\Delta m^2_{12}, -\delta, a),
\label{eq:T-CP-exact}
\end{eqnarray}
where we assumed that the matter density profile is 
symmetric about the mid-point between production 
and detection. 
Let us call this T-CP relation. 

Now we would like to get similar relations but 
keeping the sign of $\Delta m^2_{12}$ the same
because we want to find some relations among 
CP$^{(\pm)}$ and T$^{(\pm)}$ trajectories
for positive $\Delta m^2_{12}$, which is 
required by the solar neutrino data. 
By flipping the sign of $\Delta m^2_{12}$ in the RHS
of Eqs.~(\ref{eq:CP-CP-exact}) and ~(\ref{eq:T-CP-exact}), 
keeping only the oder $\Delta m^2_{12}/\Delta m^2_{13}$, 
one can get the following approximated 
CP-CP and T-CP relations~\cite{mnp1}, 
\vspace{-0.1cm}
\begin{eqnarray}
\label{CP-CP-approx}
P(\nu_{\mu} \rightarrow \nu_{e};
\Delta m^2_{13},\Delta m^2_{12}, \delta,a)
&\simeq &
P(\bar{\nu}_{\mu} \rightarrow \bar{\nu}_{e};
-\Delta m^2_{13},+\Delta m^2_{12},\pi+\delta, a) \\
P(\nu_e \rightarrow \nu_{\mu};
\Delta m^2_{13},\Delta m^2_{12}, \delta, a)
&\simeq&
P(\bar{\nu}_{\mu} \rightarrow \bar{\nu}_{e};
-\Delta m^2_{13},+\Delta m^2_{12}, \pi-\delta, a).
\label{T-CP-approx}
\end{eqnarray}
Eq.~(\ref{CP-CP-approx}) explains why the positions of 
CP$^{(+)}$ and CP$^{(-)}$ trajectories are approximately
symmetric with respect to the line $P=\bar{P}$ 
whereas Eq.~(\ref{T-CP-approx}) explains why 
the projections of CP$^{(\mp)}$ and T$^{(\pm)}$ to 
the vertical axis approximately coincide.
\vspace{-0.4cm}
\section{Problem of parameter degeneracy}
Finally, let us mention briefly about the problem 
of parameter degeneracy~\cite{BC,MNjhep01,BMW02}.
Suppose that we can measure very precisely the 
oscillation probability 
$P(\nu_\mu \rightarrow \nu_e)$ and its
CP conjugate one 
$P(\bar{\nu}_{\mu} \rightarrow \bar{\nu}_{e})$ for 
a given energy and baseline. 
This give one point $(P,\bar{P})$ in the $P-\bar{P}$ plane. 
Assuming that we know all the mixing parameters,
except for $\theta_{13}$ and $\delta$, 
there is a situation where we can find four different
CP trajectories 
which pass such a single point as illustrated in
the left panel of Fig. 5. 
This implies 
that even if we can measure the probability 
very precisely, we can not distinguish such\break
\newpage
\noindent
\hglue -0.5cm
\includegraphics[height=.33\textheight]{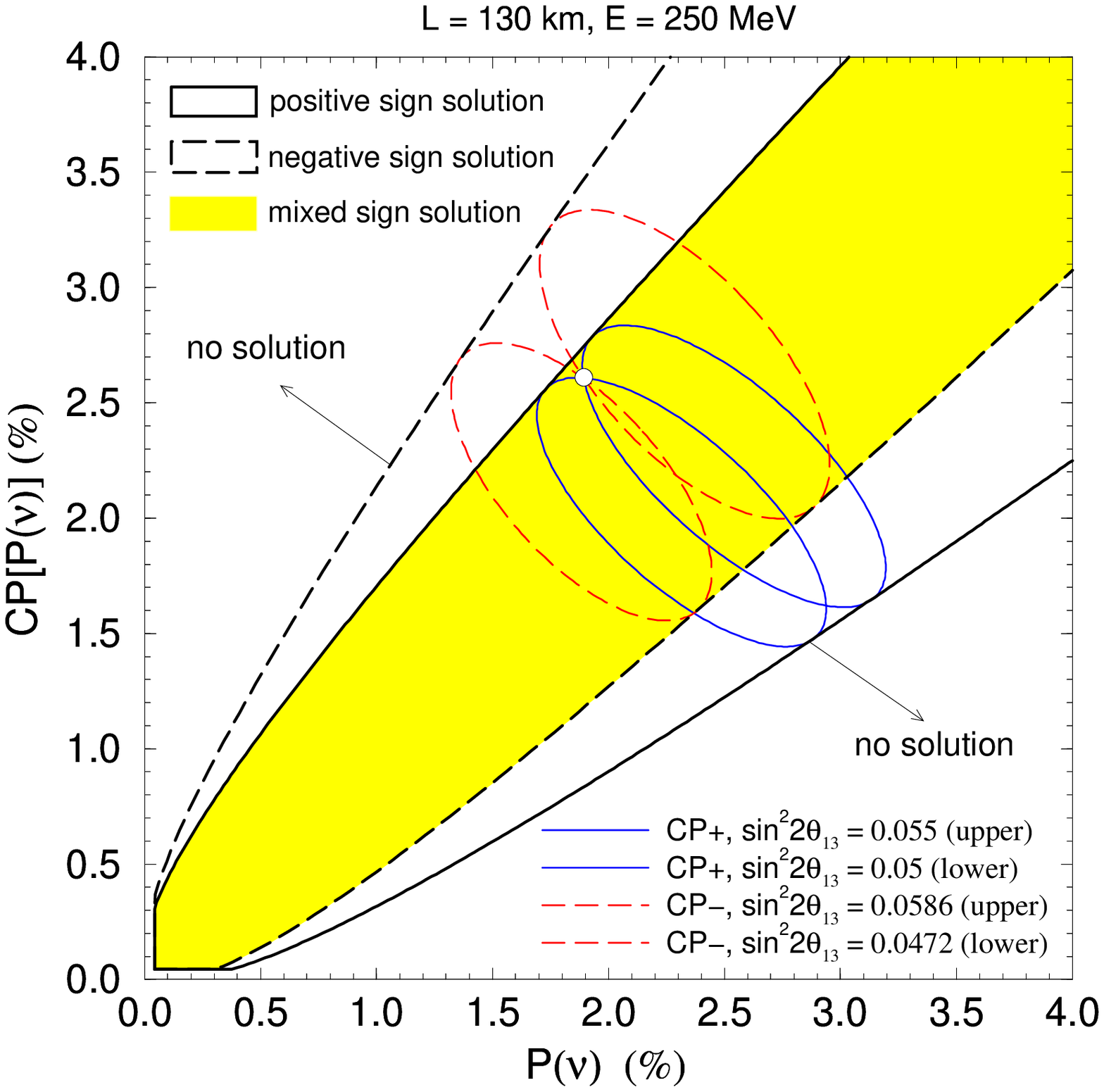}
\vglue -7.65cm
{\hskip 7.0cm}
\includegraphics[height=.33\textheight]{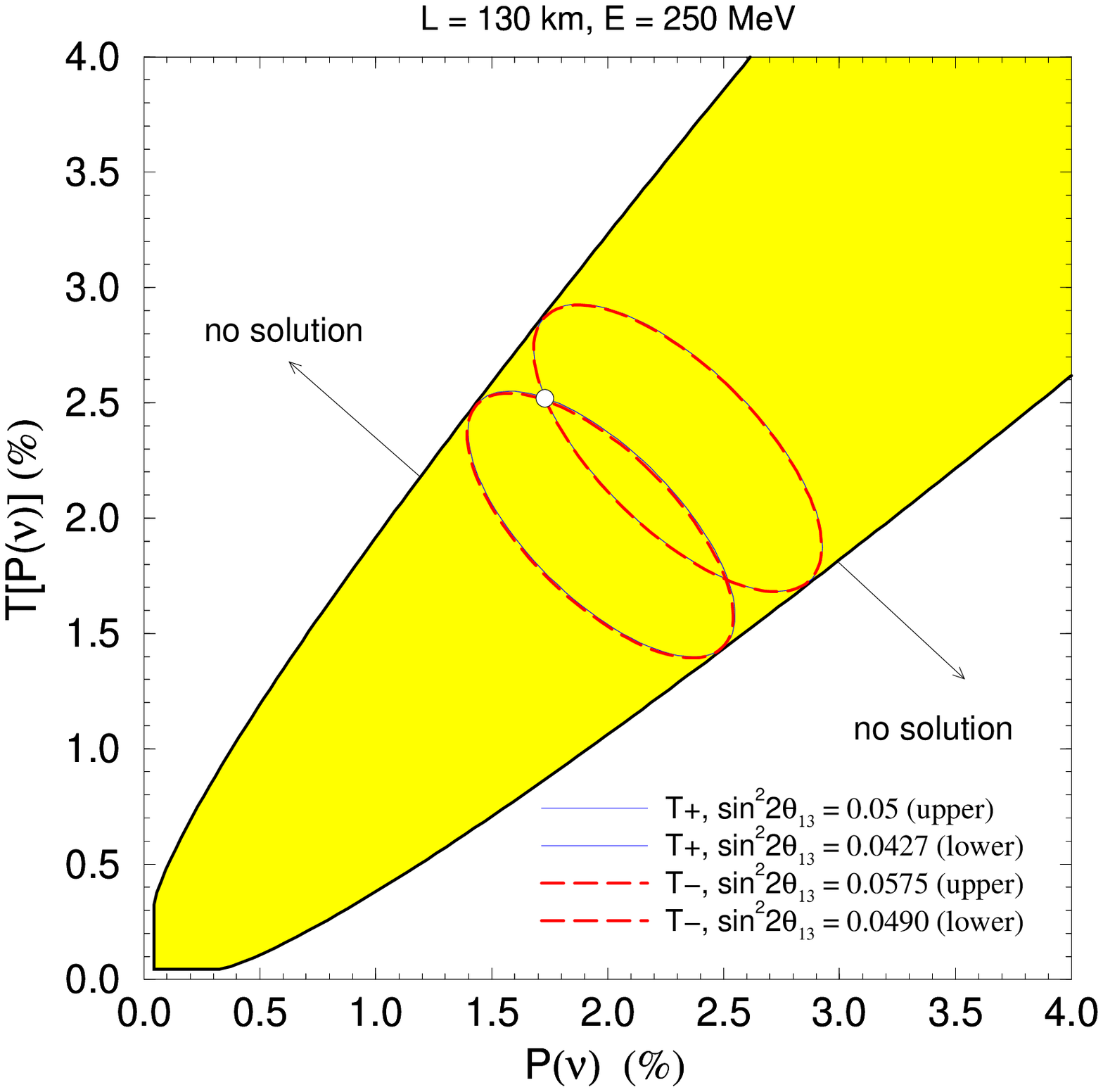}
\label{fig:dege}
\vglue -0.7cm
\noindent
{\small Fig. 5: 
Examples of the case where we have problem of 4-fold 
parameter degeneracy for CP violation (left) and 
T violation (right). Trajectories with solid (dashed) 
lines correspond to positive (negative) $\Delta m^2_{13}$
and 
$P(\nu)\equiv P(\nu_\mu \rightarrow \nu_e)$ and 
$\text{CP}[P]\equiv \bar{P}$ and $\text{T}[P]\equiv P^T.$
Same mixing parameters as in Fig. 4 except for $\theta_{13}$. 
Adopted from Ref.~\cite{mnp2}.
}
\vglue 0.3cm
\noindent
four different physical situations!
This is the essence of the problem of parameter degeneracy.
The shaded region in the plot indicate the region 
where we can not distinguish the sign of $\Delta m^2_{13}$
by just measuring one set of ($P,\bar{P}$).
In the case of T violation the problem become more serious
as there are always four trajectories 
which pass such a single point in the $P-P^T$ plane,
corresponding to four physically different cases. 
For given $P$ and $\bar{P}$ ($P^T$), possible set of 
solutions of ($\theta_{13}$, $\delta$)
can be obtained analytically~\cite{mnp2}.

How to solve this problem? 
The possible answer is to perform experiments at 
two different energies and/or two different 
baselines~\cite{BC,BMW02a,BurguetC2},
or to combine two different experiments
\cite{BurguetC2,DMM02,BMW02b,kashkari,HLW02}.
Then the degeneracy will 
be lifted and in principle, we can 
uniquely determine the oscillation parameters, 
provided that we can measure oscillation probabilities
accurately enough. See these references for detailed
discussions. 

\vspace{-0.5cm}
\section{Summary}
\vspace{-0.2cm}
We discussed some basic aspects of CP and T violation in 
neutrino oscillation in the presence of matter effect 
using bi-probability trajectory diagrams.
These trajectory diagrams are quite useful for 
qualitative understanding of the subject 
as they show the effect of CP/T violation 
as well as matter effect at the same time 
in a single plot.
We discussed the interplay among
CP/T violation and matter effects and briefly mentioned about
the problem of parameter degeneracy.
We would like to conclude that we may have a good 
chance to observe CP and T violation 
in neutrino oscillation experiments in the next few 
years, provided that both $\delta$ and $\theta_{13}$ 
are not too small.

\vspace{-0.5cm}
\begin{theacknowledgments}
\vspace{-0.3cm}
\small
HN thanks organizers of the 10 th Mexican School on Particles 
and Fields for the invitation.
HN also thanks Omar Miranda for the hospitality during his
stay at CINVESTAV in Mexico and CONACyT for 
the financial support.
\end{theacknowledgments}

%% If the bibliography is produced without BibTeX comment out the
\vspace{-0.5cm}
\bibliographystyle{aipproc}   % if natbib is available

\bibliography{sample}

\IfFileExists{\jobname.bbl}{}
 {\typeout{}
  \typeout{******************************************}
  \typeout{** Please run "bibtex \jobname" to optain}
  \typeout{** the bibliography and then re-run LaTeX}
  \typeout{** twice to fix the references!}
  \typeout{******************************************}
  \typeout{}
 }

\end{document}